\documentclass[12pt]{article}
\usepackage{graphicx}
\usepackage{cite}
\newcommand{\xrm}[1]{\mbox{ #1}}
\newcommand{\fnd}[2]{\frac{\textstyle #1}{\textstyle #2}}
\newcommand{\icgp}[1]{\resizebox{12cm}{!}{\includegraphics{#1}}}
\newcommand{\eqr}[1]{Eq.~(\ref{#1})}
\newcommand{\dss}{$D^*_{\!s0}$}
\newcommand{\ds}{$D^*_{\!s0}$(2317)}
\newcommand{\dz}{$D_0^*$(2300--2400)}
\newcommand{\dzs}{$D_0^*$}
\newcommand{\chc}{$\chi_{c0}$}
\newcommand{\chcp}{$\chi'_{c0}$}
\newcommand{\kzs}{$K_0^*$}
\newcommand{\kz}{$K_0^*$(1430)}
\newcommand{\kzp}{$K_0^*$(1820)}
\newcommand{\ka}{$K_0^*$(800)}
\newcommand{\si}{$f_0$(600)}
\begin{document}
\title{From the $\kappa$ via the $D^*_{\!s0}$(2317) to the $\chi_{c0}$:\\ 
connecting light and heavy scalar mesons}
\author{
Eef van Beveren\\
{\normalsize\it Centro de F\'{\i}sica Te\'{o}rica}\\
{\normalsize\it Departamento de F\'{\i}sica, Universidade de Coimbra}\\
{\normalsize\it P-3000 Coimbra, Portugal}\\
{\small http://cft.fis.uc.pt/eef}\\ [.3cm]
\and
Jo\~{a}o E.\ G.\ N.\ Costa, Frieder Kleefeld, George Rupp\\
{\normalsize\it Centro de F\'{\i}sica das Interac\c{c}\~{o}es Fundamentais}\\
{\normalsize\it Instituto Superior T\'{e}cnico, Edif\'{\i}cio Ci\^{e}ncia}\\
{\normalsize\it P-1049-001 Lisboa Codex, Portugal}\\
{\small jegnc2@hotmail.com, kleefeld@cfif.ist.utl.pt,
george@ajax.ist.utl.pt}\\ [.3cm]
{\small PACS number(s): 14.40.Ev, 14.40.Lb, 14.40.Gx, 13.25.-k}\\ [.3cm]
}

\maketitle

\begin{abstract}
Pole trajectories connecting light and heavy scalar mesons, both broad
resonances and quasi-bound states, are computed employing a simple
coupled-channel model. Instead of varying the coupling constant as in
previous work, quark and meson masses are continuously changed, so as
to have one scalar meson evolve smoothly into another with different
flavor(s). In particular, it is shown, among several other cases, how the
still controversial $K_0^*$(800) turns into the established $\chi_{c0}$,
via the disputed $D_s$(2317). Moreover, a $\chi'_{c0}$(3946) is predicted,
which may correspond to the recently observed $Y$(3943) resonance. These
results lend further support to our unified dynamical picture of all scalar
mesons, as unitarized $q\bar{q}$ states with important two-meson components.
\end{abstract}
\clearpage

After more than four decades, understanding the scalar mesons continues to
pose serious difficulties to theorists as well as experimentalists. Still
today, no consensus exists about the lightest and oldest structures in the
scalar-meson sector, namely the $\sigma$ (\si\ \cite{PDG04}) \cite{AAABA89_04}
and the $\kappa$ (\ka\ \cite{PDG04}) \cite{ABA02_05,AL05}.
But also the discovery of the surprisingly light charmed scalar \ds\
\cite{ABK03}, though  giving a new boost to meson spectroscopy in general,
has not contributed to the understanding of scalar mesons, as can be seen
from the many different approaches to the \ds\ in the literature (see
Ref.~\cite{ds2317} for a representative, albeit not totally exhaustive, list
of references). Here, we shall focus on a formalism which successfully
describes all mesonic resonances, including the scalar mesons.

In Ref.~\cite{BR03} it was shown that the \ds\ meson can be straightforwardly
explained as a normal $c\bar{s}$ state, but strongly coupled to the nearby $DK$
channel, which is responsible for its low mass. The framework for this
calculation was a simple coupled-channel model, which had been employed
previsously \cite{BR01}
to fit the $S$-wave $K\pi$ phase shifts, and predict the now listed\cite{PDG04}
\ka, besides reproducing the established \kz. Furthermore, another charmed
scalar meson was predicted in Ref.~\cite{BR03}, i.e., a broad $D_0^*$ resonance
above the $D\pi$ threshold, somewhere in the energy region 2.1--2.3 GeV, which
may correspond to the $D_0^*$(2300--2400) \cite{AL03,PDG04}. Also higher-mass
$D^*_{\!s0}$ and $D^*_0$ resonances were foreseen \cite{BR03}, which have not
been observed so far.

The purpose of this Letter is to show the interconnection of the scalar mesons
\ka, \dz, \ds\ with one another, and also with the established
$\chi_{c0}$(3415) \cite{PDG04}. Moreover, the same interconnection will be
demonstrated for the higher-mass recurrences of these scalars, thereby finding
a candidate for the very recently observed $Y$(3943) charmonium state
\cite{A05}. For that
purpose, we shall employ the above-mentioned coupled-channel model, but now for
fixed, physical coupling, while quark and threshold masses will be varied.
Thus, a continuous and smooth transition can be achieved from one scalar meson
to another. Crucial here will be a mass scaling \cite{BR04a,K04} of the two
parameters modeling the off-diagonal potential that couples the confined and
decay channels. This way, these two parameters, identical to the ones used in
Refs.~\cite{BR01,BR03,BR04a}, suffice to reasonably describe a vast range of
distinct scalar mesons. On the other hand, the confinement and 
quark-mass parameters are taken at their usual published values.

Starting point is a simple, intuitive coupled-channel model, describing a
confined $q\bar{q}$ system, coupled to one meson-meson channel accounting
for the possibility of real or virtual decay via the $^3\!P_0$ mechanism. If
the transition potential is taken to be a spherical delta function, the
$1\times1$ inverse $K$ matrix can be solved in closed form, reading \cite{BR01}
\begin{equation}
\cot\left(\delta_{\ell}(p)\right) =
\fnd{n_{\ell}(pa)}{j_{\ell}(pa)}\,-\,
\left[ 2\lambda^{2}\!\mu\,pa j^{2}_{\ell}(pa)
\sum_{n=0}^{\infty}\fnd{B_{n\ell_{c}}}{E-E_{n\ell_{c}}}\right]^{-1} ,
\label{deltaBW}
\end{equation}
where $j_\ell,n_\ell$ are spherical Bessel and Neumann functions, respectively,
$\lambda$ is the $^3\!P_0$ coupling, $a$ is the delta-shell radius,
$E_{n\ell_c}$ are the energies of the bare confinement spectrum, $B_{n\ell_c}$
are the corresponding weight factors, $p$ is the on-shell relative momentum in
the two-meson channel, given by the kinematically relativistic expression
\begin{equation}
4s\,p^{2} = \left[s-(M_1+M_2)^2\right]\left[s-(M_1-M_2)^2\right] \; ,
\label{prel}
\end{equation}
and $\mu$ is the ensuing relativistic reduced mass
\begin{equation}
\mu \, \equiv \, \frac{1}{2}\fnd{dp^{2}}{d\sqrt{s}} \, = \, \fnd{\sqrt{s}}{4}
\;\left[ 1 - \left(\fnd{M_1^{2}-M_2^{2}}{s}\right)^{2}\right]
\; .
\label{mu}
\end{equation}
As the present paper deals with scalar mesons, we have $\ell\!=\!0$ and
$\ell_c\!=\!1$ in \eqr{deltaBW}. Moreover, since only ground states and first
radial excitations are considered here, we shall approximate the infinite
sum in \eqr{deltaBW} by two confinement-spectrum states plus one rest term,
also sticking to the numerical values used in Refs.~\cite{BR01,BR03,BR04a},
namely $B_{01}\!=\!1.0$, $B_{11}\!=\!0.2$, and $B_{21}\!=\!E_{21}\!=\!\infty$,
with $B_{21}/E_{21}\!=\!1$. As for the two confinement levels, we parametrize
them by a harmonic oscillator \cite{BR03,BR04a}, i.e.,
\begin{equation}
E_{n1} \; = \; (2n+2.5)\,\omega+m_{q_1}+m_{q_2} \; ,
\label{HO}
\end{equation}
where $\omega\!=\!0.190$ GeV, $m_n\!=\!0.406$ GeV ($n\!=\!u,d$),
$m_s\!=\!0.508$ GeV, and $m_c\!=\!1.562$ GeV, as in previous work
\cite{BRRD83,BR01,BR03,BR04a,BR04b}. Finally, we assume a mass scaling of the
parameters $a$ and $\lambda$ given by \cite{BR04a,K04}
\begin{equation}
a_{ij}\,\sqrt{\mu_{ij}}\; =\;\xrm{constant}
\;\;\;\;,\;\;\;\;
\lambda_{ij}\,\sqrt{\mu_{ij}}\; =\;\xrm{constant}
\; ,
\label{scaling}
\end{equation}
where the labels $ij$ refer to a particular combination of quark flavors,
and $\mu_{ij}\equiv m_{q_i}m_{q_j}/(m_{q_i}\!\!+\!m_{q_j})$ is the corresponding
reduced quark mass.  This procedure ensures flavor invariance of our equations.
Using then the values $\lambda_{ns}\!=\!0.75$ GeV$^{-3/2}$ and $a_{ns}\!=\!3.2$
GeV$^{-1}$ from the fit to the $K\pi$ $S$-wave phase shifts in
Ref.~\cite{BR01}, we have fixed all our parameters,\footnote
{Note that we use here somewhat shifted confinement levels as compared to
Ref.~\cite{BR01}, namely the ones following from \eqr{HO}. This gives rise
to a slightly lighter and broader $\kappa$ meson, and a heavier \kz.}
which allows to show the
predic\-tive power of our approach. For the required input mesons masses, we
take the isospin-averaged values \cite{PDG04} $M_\pi\!=$ $0.1373$ GeV,
$M_K\!=\!0.4957$ GeV, and $M_D\!=\!1.867$ GeV.

Now we can compute pole trajectories in the complex
energy plane for scalar resonances and (virtual) bound states, by searching
the values of $s$ for which $\cot\delta_0\!\left(p(s)\right)=i$.
However, instead of freely varying $\lambda$ as in previous work, we shall
keep $\lambda_{ns}$ fixed at its physical value of 0.75 GeV$^{-3/2}$, while
changing instead one of the quark masses, as well as one of the meson masses 
in the decay channel. This way we can make one scalar meson turn into another.
For instance, by letting
\begin{equation}
\left. \begin{array}{lcl}
m_{q_1} & = & m_n + \alpha\,(m_c-m_n) \\
M_1     & = & M_\pi - \alpha\,(M_D-M_\pi)
\end{array} \right\} \; , \;\; 0 \leq \alpha \leq 1 \; ,
\label{mvar}
\end{equation}
we smoothly change the $\kappa$ ($n\bar{s}$) meson, coupling to the $\pi K$
channel, into the \ds\ ($c\bar{s}$), coupling to $DK$. The poles themselves
are numerically found and checked with two independent methods, i.e., the \em
MINUIT \em \/package of CERN \cite{CERN94}, and \em MATHEMATICA \em
\/\cite{W03}.

\begin{figure}[htbp]
\begin{center}
\icgp{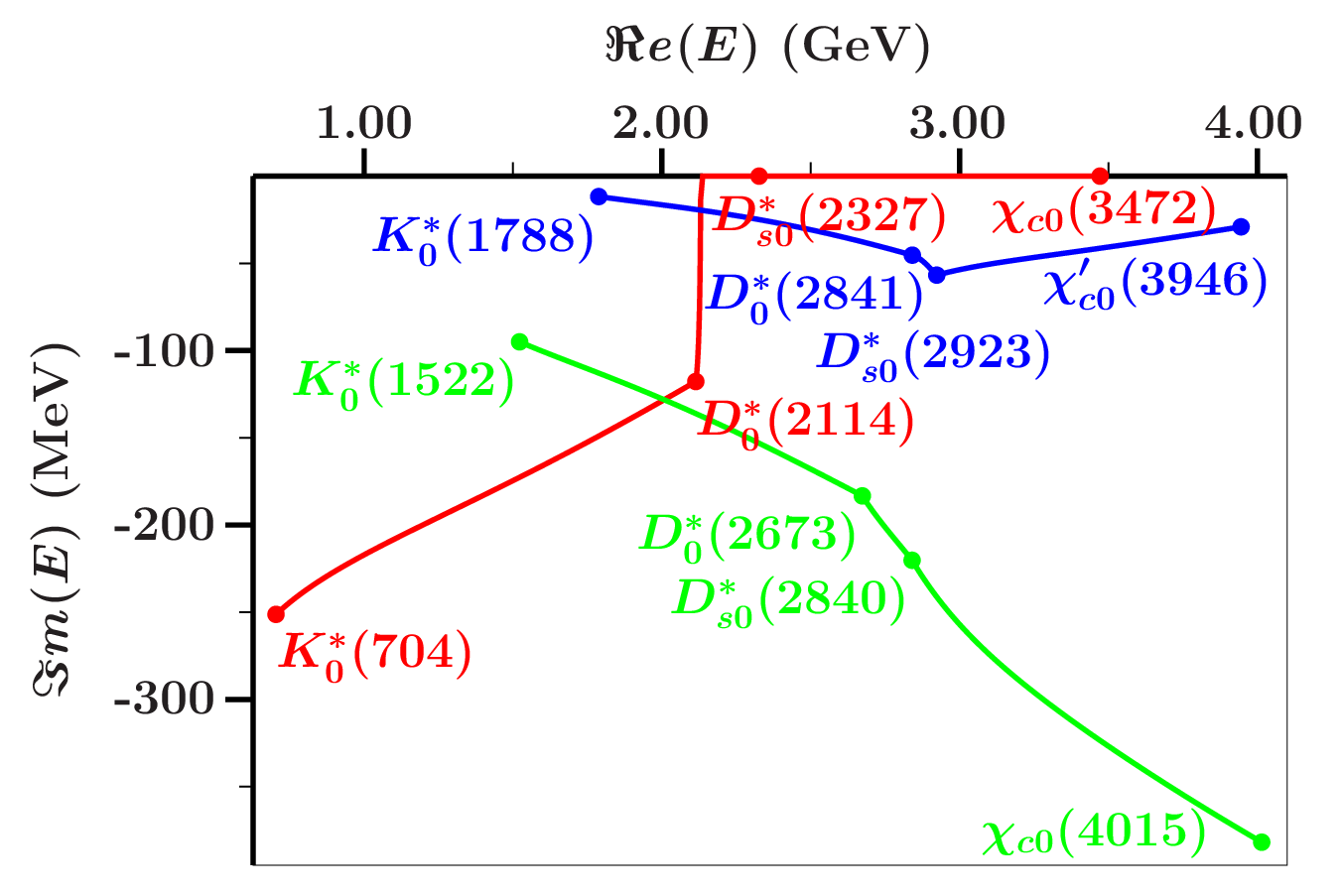}
\end{center}
\caption[]{Scalar-meson pole trajectories in the complex energy plane. Dots
represent predicted resonances or bound states. See text and 
Eqs.~(\ref{trareal},\ref{traimag}) for further details.}
\label{alltraj}
\end{figure}

In Fig.~\ref{alltraj}, one sees in one glimpse the nine trajectories

\begin{equation}
\begin{array}{l}
\mbox{a:}\;\;K_0^*(704)\to D_0^*(2114)\to D^*_{s0}(2327)\to\chi_{c0}(3472),\\
\mbox{b:}\;\;K_0^*(1522)\to D_0^*(2673)\to D^*_{s0}(2840)\to\chi_{c0}(4015),\\
\mbox{c:}\;\;K_0^*(1788)\to D_0^*(2841)\to D^*_{s0}(2923)\to\chi'_{c0}(3946),
\end{array}
\label{trareal}
\end{equation}
where the numbers between parentheses are the real parts (in MeVs) of the
respective resonance/bound-state poles, the corresponding imaginary parts being
\begin{equation}
\begin{array}{l}
\mbox{a:}\;\;K_0^*(-251)\to D_0^*(-118)\to D^*_{s0}(0)\to \chi_{c0}(0),\\
\mbox{b:}\;\;K_0^*(-95)\to D_0^*(-183)\to D^*_{s0}(-220)\to \chi_{c0}(-382),\\
\mbox{c:}\;\;K_0^*(-12)\to D_0^*(-45)\to D^*_{s0}(-57)\to \chi'_{c0}(-29).
\end{array}
\label{traimag}
\end{equation}
Before discussing the actual trajectories, a few remarks are due concerning
the precise values found for the pole positions. Clearly, for such a simple
model without any fitting freedom, moreover covering a vast energy range, a
very accurate reproduction of the masses and widths of all experimentally
observed mesons cannot, and should not even be expected. In particular, the
inclusion of only the lowest, dominant decay channel for each state will
certainly reflect itself in one way or another. For instance, the much too
small width of our \kzs(1788), which should correspond to the observed
\cite{AS97} \kzp, is probably owing to the neglect of the important $K\eta'$
channel. Furthermore, the somewhat too large mass of our \chc(3472), as
compared to the established \cite{PDG04} \chc(3415), may very well be due
to the omission of vector-vector decay channels, which are relevant for
charmonium ground states \cite{BDR80}. Note, however, that the latter
discrepancy of 57 MeV is quite insignificant when compared to the huge
coupled-channel shifts in charmonium recently found in Refs.~\cite{B05,K05}.
Notwithstanding, a clear identification can be made of our broad
\kzs(704), \dzs(2114), and \kzs(1522) states with the listed \cite{PDG04}
\ka, \dz, and \kz\ resonances, respectively. Here, one should also notice that
we give the real parts of the pole positions of our resonances, which usually
do not coincide with the experimental masses resulting from Breit-Wigner fits
when the widths are large.
As for the remaining observed mesons, our \dss(2327) is very close to the \ds,
while our \chcp(3946), with a width of about 60 MeV, seems a good candidate
for the brand-new \cite{A05} charmonium state $Y$(3943). Finally, we predict
the two medium-broad charmed mesons \dzs(2841) and \dss(2923), so far
undetected, as well as the very broad states \dzs(2673), \dss(2840), and
\chc(4015), which will be extremely hard to observe at all. In any case, the
predictions for the latter higher-mass states may change significantly when 
additional decay channels are taken into account. 

\begin{figure}[htbp]
\begin{center}
\icgp{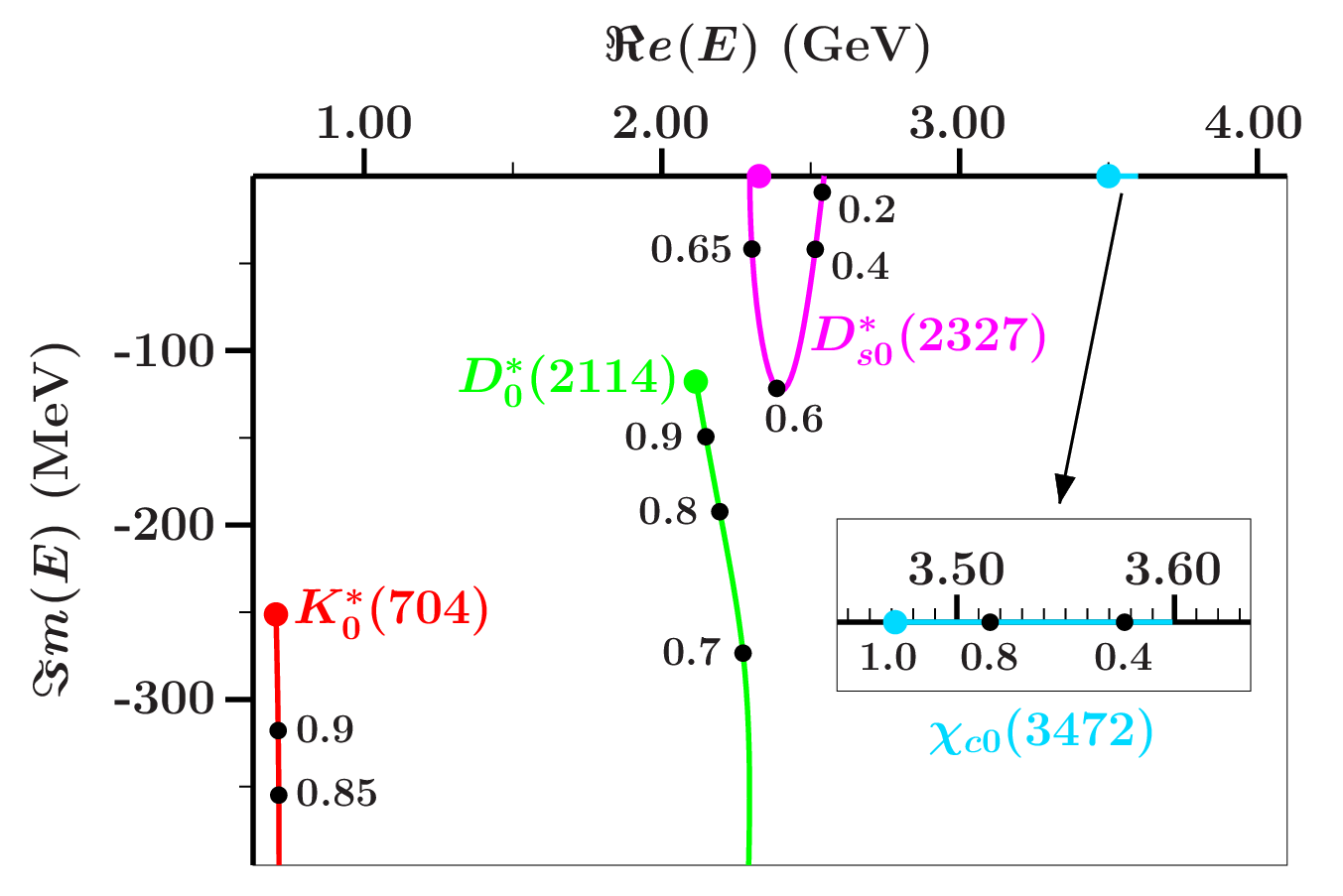}
\end{center}
\caption[]{Pole trajectories of lowest states as a function of $\lambda$.
Numbers indicate reductions relative to the maximum value.}
\label{tracpl}
\end{figure}

Turning now to the trajectories themselves, it is remarkable to observe that
physical states with radically disparate widths can be continuously connected
to one another in flavor. This is one of the reasons why scalar-meson
spectroscopy is so intricate. Moreover, as we shall see below, states on the
same mass trajectory can have different origins when viewed as $q\bar{q}$
states distorted by meson loops, which point will become clearer when we study
Fig.~\ref{tracpl}. Anyway, the first radial excitations of the $n\bar{s}$,
$c\bar{n}$, $c\bar{s}$, and $c\bar{c}$ systems are all on the same trajectory
in Fig.~\ref{alltraj}, i.e., the one connecting the \kzs(1788) and \chcp(3946).

In Fig.~\ref{tracpl}, the lowest states for the various flavor combinations
are displayed again, but now also showing how the corresponding poles move
when the coupling $\lambda$ is reduced from its fixed value. We see that the
\kzs(704) and the \dzs(2114) appear to find their origin in the continuum,
corresponding to infinitely negative imaginary parts of their pole positions,
while the \dss(2327) and \chc(3472) are connected to the confinement spectrum,
with poles on the real axis.
This is quite surprising for the nearby pair \dzs(2114)--\dss(2327). However,
even the physical \ds\ itself can be either interpreted as a ``confinement''
state \cite{BR04a,RKB05}, or a ``continuum'' state \cite{BR03}, depending on
tiny changes in e.g.\ the parameter $a$. What this figure also shows is 
an extremely delicate balance of coupling effects. With a small decrease
of $\lambda$, the \dz\ and especially the $\kappa$ meson would become even
broader and thus almost impossible to observe experimentally, while the
\ds\ would be a resonance or a virtual state instead of a quasi-bound state.

\begin{figure}[htbp]
\begin{center}
\icgp{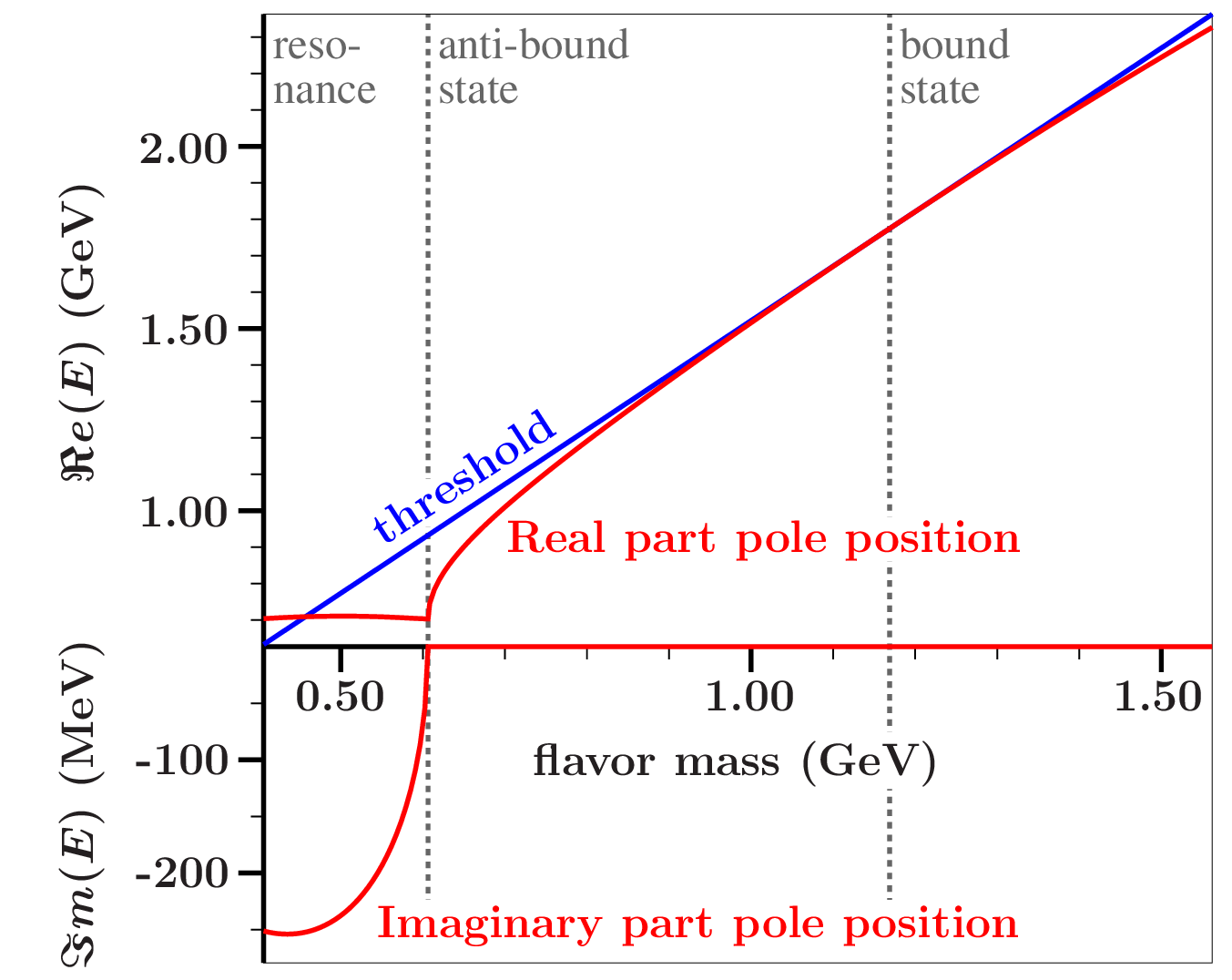}
\end{center}
\caption[]{Real and inaginary parts of \kzs(704) pole turning into the
\dss(2327), as a function of varying quark mass. Straight dashed line stands
for decay threshold (real).}
\label{KDs}
\end{figure}

Finally, in Fig.~\ref{KDs} a direct transition of the \kzs(704) into the
\dzs(2327) is displayed, by letting $m_n\to m_c$, $M_\pi\to M_D$ as in
\eqr{mvar}, and moreover in a different fashion. Namely, instead of giving
the pole positions in the complex energy plane, we now plot the corresponding
real and imaginary parts as a function of the varying quark mass, as well as
the proportionally changing threshold value. It is striking to see how the
\kzs(704) resonance quickly turns into a virtual bound state, while its real
part remains almost constant. Here, we probably see the kinematical Adler zero
\cite{RKB05} at work, which rapidly moves away as one of the decay masses
increases from $M_\pi$, thus allowing the pole to approach the real axis.
Then, the pole moves along the real axis as a virtual state, until it touches
the threshold at about 1.76 GeV, after which it becomes a bound state. Notice
again the tiny margin, at least on this scale, by which the \dss(2327) is
bound.


To conclude, in the present paper we have shown how several light and heavy
scalar mesons can be linked to one another, by continuously varying some of
the involved flavor and decay masses. This way, the common dynamical nature of
the studied --- and probably all --- scalar mesons, as ordinary $q\bar{q}$
states but strongly distorted due to coupled channels, is further
substantiated.  Thus, tetraquarks and other exotic configurations are not
needed in this context. Moreover, we deduce that labeling
scalar mesons as $q\bar{q}$ states as \em opposed \em \/to dynamical
meson-meson resonances makes no sense, in view of the tiny parameter
variations needed to turn one kind of pole into another. Rather, scalar
mesons should be considered nonperturbatively dressed $q\bar{q}$ systems,
with large meson-meson components, no matter if one uses a coupled-channel
quark model \cite{BRMDRR86} or e.g.\ the quark-level linear sigma model
\cite{DS95}. As a consequence, the spectroscopy of scalar mesons is much
more complex than for ordinary mesons, with the total number of potentially
observable states being different from the number of confined, bare $q\bar{q}$
states.

In the course of this analysis, we have also found a candidate for the new
charmonium state $Y$(3943) \cite{A05}. It is true that such a resonance, if
indeed a scalar,
should dominantly decay to $D\bar{D}$, a mode which has not been observed yet.
However, the reported decay $Y(3943)\to \omega J/\Psi$ is OZI-forbidden, so
that it cannot account for the measured sizable width of
$\Gamma=87\,(\pm22\pm26)$ MeV.

We thank D.~V.~Bugg for enlightening discussions about scalar mesons in
general, and the new charmonium state $Y$(3943) in particular.
This work was supported in part by the {\it Funda\c{c}\~{a}o para a
Ci\^{e}ncia e a Tecnologia} \/of the {\it Minist\'{e}rio da Ci\^{e}ncia,
Tecnologia e Ensino Superior} \/of Portugal, under contract
POCTI/FP/FNU/50328/2003 and grant SFRH/BPD/9480/2002.

\end{document}